\documentclass[prd,singlecolumn,superscriptaddress,showpacks]{revtex4}
\usepackage[utf8]{inputenc}
\usepackage{empheq}
\usepackage{amsmath}
\usepackage{amssymb}
\usepackage{stmaryrd}
\usepackage{dsfont}
\usepackage{graphicx}
\usepackage{stackrel}
\usepackage{color}
\usepackage{comment}
\usepackage{hyperref}

\def \ee{\end{equation}}
\def \be{\begin{equation}}
\def \eea{\end{eqnarray}}
\def \bea{\begin{eqnarray}}

\begin{document}

\title{
Almost
relevant corrections for direct measurements of electron's g-factor}

\author{Benjamin Koch}
\email{bkoch@fis.puc.cl}
\affiliation{Institut f\"ur Theoretische Physik,
 Technische Universit\"at Wien,
 Wiedner Hauptstrasse 8-10,
 A-1040 Vienna, Austria}
 \affiliation{Pontificia Universidad Cat\'olica de Chile \\ Instituto de F\'isica, Pontificia Universidad Cat\'olica de Chile, \\
Casilla 306, Santiago, Chile}

\author{Felipe Asenjo}
\affiliation{Facultad de Ingenier\'ia y Ciencias, Universidad Adolfo Ib\'a\~nez, Santiago 7491169, Chile.}

\author{Sergio Hojman}
\affiliation{Departamento de Ciencias, Facultad de Artes Liberales,
Universidad Adolfo Ib\'a\~nez, Santiago 7491169, Chile}
\affiliation{Departamento de F\'isica, Facultad de Ciencias, Universidad de Chile, Santiago 7800003, Chile.}
\affiliation{Centro de Recursos Educativos Avanzados, CREA, Santiago 7500018, Chile.}

\begin{abstract}
We revisit the observable used for the direct measurements of the electron's $g$ factor. 
This is done by
considering the sub-leading effects of the large magnetic background field and virtual Standard Model processes.
We find substantial corrections to the Landau levels of the electron.
Implications for the observed magnetic moment and the tension between direct and indirect measurement are discussed.\\

Youtube: \href{https://www.youtube.com/watch?v=Yr08HgEbZIQ}{paperpitch}

\end{abstract}

\maketitle

\tableofcontents

\section{Introduction}
\label{sec_Intro}

\subsection{Electron $g-2$ in a nut-shell}

Quantum electrodynamics (QED) is
one of the cornerstones of our understanding of the electron, its properties, and interactions \cite{Feynman:1949zx,Feynman:1950ir}.
This theory allows  performing the interplay between theoretical prediction and experimental confirmation to unprecedented precision.
The anomalous magnetic moment of the
electron plays a pivotal role in this ongoing success. 

Luttinger and Schwinger made the first theoretical calculations suggesting that
 the magnetic moment of the electron should differ from two~\cite{Schwinger:1948iu,Luttinger:1948zz}. Over the decades, these calculations became more and more precise     \cite{Aoyama:2007mn,Aoyama:2012wj,Aoyama:2017uqe}, such that they currently include even tenth order in perturbation theory (for a review see  \cite{Aoyama:2019ryr}).
 For recent developments concerning the anomalous magnetic moment of the muon see \cite{Aoyama:2020ynm}.
 
 Such an amazing effort on the theoretical side only makes sense, if the experimental observations have a precision that is comparable to the theoretical uncertainty. Since the first experimental observation of $g\neq 2$~\cite{Foley:1948zz,Kusch:1948mvb}, the measurements have been continuously improved. Currently the most precise direct measurement of the electron's $g$ factor go down to 13 significant digits~\cite{Hanneke:2010au,Hanneke:2008tm,Gabrielse:2006gg}. These experiments, which will be discussed in more detail below, use Penning traps.
 A comparison between theory and experiment at such precision does not only allow for better testing of QED, but it also is an excellent tool to test the Standard Model of particle physics (SM) and to
 to constrain models that go beyond the Standard Model of particle physics (BSM)~     \cite{Broggio:2014mna,Essig:2013vha,Endo:2012hp,Giudice:2012ms,Kahn:2016vjr,Raggi:2015yfk}.

These direct measurements can also be complemented by  measurements of the fine structure constant $\alpha$, which, by virtue of the theoretical predictions, allow  inferring the value of $g$   \cite{Bouchendira:2010es,AmelinoCamelia:2010me}.
Three years ago, such a comparison
showed a significant tension
of $2.4$ standard deviations between the directly measured and the inferred magnetic moment~ \cite{Parker:2018vye}.
This triggered new hope for 
the search of physics beyond the Standard Model. Currently, the most prominent examples of these models, which are capable of resolving the tension, are~\cite{AmelinoCamelia:2008qg,Crivellin:2018qmi,Davoudiasl:2018fbb,Badziak:2019gaf,Bauer:2019gfk,Endo:2019bcj,Liu:2018xkx,Hiller:2019mou,Han:2018znu,Dutta:2018fge}.
However, last year a new experiment
gave a discrepancy pointing in the opposite direction,
but with a smaller absolute value, such that 
one can not speak of a significant tension any more~\cite{Morel:2020dww}.

In this paper, we revisit the observable which is used
to determine the electron's  anomalous magnetic moment in the most precise direct measurements~\cite{Hanneke:2010au,Hanneke:2008tm,Gabrielse:2006gg}.
The paper is organized as follows. 
In the following subsection \ref{subsec_setting}, we give an idealized and thus simplified summary of the
key observable. We then motivate the study by 
discussing and comparing the orders of magnitude involved in this problem \ref{subsec_expect}. 
Section \ref{sec_Quant} uses an effective Lagrangian approach. It contains four parts in which the 
path from Lagrangian corrections to changes in the observable is described. In section \ref{sec_Discussion},
these findings are then applied and discussed in the context of a comparison between direct and indirect measurements of $(g-2)$.
A summary and conclusion is given in section \ref{sec_Concl}.
While preparing this paper, we noticed the preprint \cite{Kim:2021ydg}, which also 
considers the effect of a finite magnetic field. It is, however, mainly discussing the implications for the anomalous magnetic moment of the muon.

\subsection{Idealized experimental setting}
\label{subsec_setting}

The motion of an electron 
in a real Penning trap has four eigenfrequencies, which are the spin-, cyclotron-, axial-, and magnetron-frequency.
For most parts of this paper, it is, however, sufficient to consider the idealized case, where electric field and magnetron effects are absent.

The magnetic field strength, used in this experiment,
can be inferred from the measured cyclotron frequency of about $\nu_c=\omega/(2 \pi)\approx 149~$GHz~\cite{Hanneke:2010au}
\be\label{BB}
|B|=\frac{2\pi \nu_c m}{e_0}\approx 5.3~T.
\ee
Here, $e_0$ is the absolute value of the electron charge.
The energy-eigenvalues of this
system are~\cite{Brown:1985rh}
\be\label{EnergiesOdom}
E_n^\pm=\frac{1}{2}
\left( 2 n+ 1\pm 1 \right)h \nu_c
\pm \frac{a}{2}h \nu_c
-\frac{1}{8}
\frac{h^2 \nu_c^2}{m c^2} \left( 2 n+ 1\pm 1 \right)^2,
\ee
where the last term
is the leading relativistic correction.
The parameter of interest in (\ref{EnergiesOdom}) is
the anomalous magnetic moment 
\be
a\equiv \frac{(g-2)}{2}.
\ee 
Measurements of transitions between these
energies are used to determine the value of  $a$.
It turns out that a particular ratio
 of such transition energies cancels
several effects of imperfections and is thus particularly suited for this purpose~\cite{Hanneke:2010au,Hanneke:2008tm,Gabrielse:2006gg}
\be\label{aFormula}
a_d=\frac{E_0^+-E_1^-}{E_1^+-E_0^++3 h \delta /2}.
\ee
The subscript of $a_d$ refers to 
``direct'' measurement.

\subsection{Corrections to expect}
\label{subsec_expect}

\begin{figure}[ht!]
\begin{center}
\includegraphics[width=1.0\columnwidth]{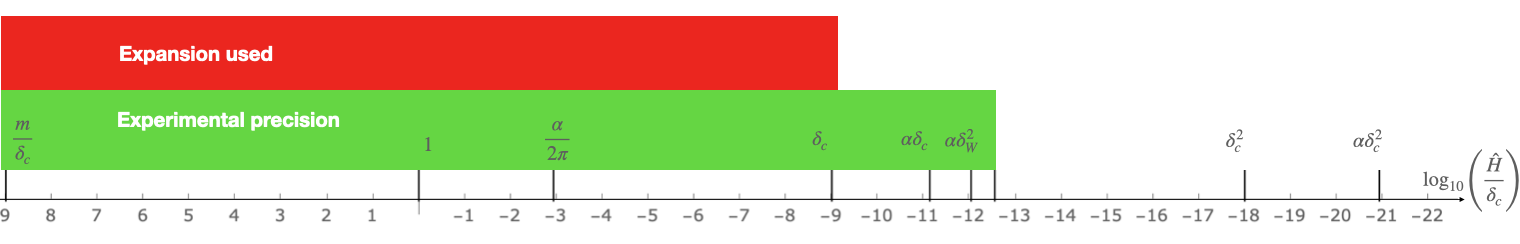} 
\end{center}
\caption{
Orders of magnitude of the expansion parameters (\ref{deltac}, \ref{deltaalpha}, \ref{deltaW}). 
The precision reached by current experiments is indicated by the green bar. The theoretical expansion order used to
interpret the experimental results~(\ref{EnergiesOdom}, \ref{aFormula}) is indicated by the red bar.
The energies in this sketch are normalised to the dimensionless ratio $\sim (\Delta E/(m c^2 \delta_c) )$.
}
\label{fig:Moti}
\end{figure}

Quantum field theoretical descriptions are effective many-particle theories, that consider excitations of all fields $(\psi, A_\mu, \dots)$.
This leads to corrections to the quantum mechanical Dirac equation
\be
i \hbar \psi= H_D \psi.
\ee
When discussing these corrections for the above experimental setting we have three
small dimensionless numbers at our disposal,
which can be used to obtain good estimates.
In order to compare these quantities, we introduce a common small expansion parameter $\epsilon$.
\begin{itemize}
\item
First, for the non-vanishing magnetic field  in~(\ref{BB}), there is
\be\label{deltac}
\delta_c \equiv\frac{h \nu_c}{m c^2}\sim 10^{-9}\sim \epsilon^{ 4}.
\ee
The Foldy-Wouthuysen transformation~\cite{Foldy:1949wa}
is  essentially based on an expansion in (\ref{deltac}) since
relativistic energy corrections to the rest mass are given as series of the form $1 + \delta_c+ \delta_c^2 + \delta_c^3\dots$.
\item 
Second, there is the fine-structure constant
\be \label{deltaalpha}
 \alpha\sim 2\pi a  \sim 0.007\sim \epsilon.
\ee
Perturbation theory in QED is an expansion in (\ref{deltaalpha}). The famous anomalous magnetic moment contributes to the energies with a term $\alpha \delta_c$.
\item
The third dimensionless parameter is the ratio between the electron mass $m$ and
and the mass of the $W$-boson
\be\label{deltaW}
\delta_W= \frac{m}{m_W}\sim 10^{-5}\sim \epsilon^{2.5}.
\ee
The leading contribution to the Hamiltonian of these corrections appears at quadratic order in $\delta_W$ and is induced by loop corrections. It is suppressed by $\alpha \delta_c \delta_W^2$ or by $\alpha \delta_W^2$, as shown in (\ref{WeakSup}).
\end{itemize}
The above expansions are present in energy formulas like (\ref{EnergiesOdom}). To see the relevance of the different terms in this expansion it is instructive to compare them to the experimental precision of~\cite{Hanneke:2008tm}. In this experiment ``$a$'' was determined by measuring a dimensionless ratio $\sim (\Delta E/(m c^2 \delta_c) )$ between different transition energies ($\Delta E, \delta_c$) of the quantum mechanical system to a precision of $2.8 \cdot 10^{-13}$.
This comparison of magnitudes is shown in figure \ref{fig:Moti}.
Remembering that the interpretation of the experimental measurement considers
dimensionless corrections up to order $\sim \delta_c$,
 figure \ref{fig:Moti} gives some expected and some surprising insight.
One sees that sub-sub-leading order relativistic corrections ($\sim \delta_c^2$), or loop corrections with three external photon lines ($\alpha \delta_c^2$) are well beyond experimental reach.
On the other hand, figure  \ref{fig:Moti} shows that
loop corrections with two external photon lines $\sim \alpha \delta_c$ and loop corrections involving a weak interaction and a single external photon line $\sim \alpha \delta_W$ can be expected to give
relevant corrections to the interpretation of experimental results.
From the point of view of a combined expansion in (\ref{deltac}, \ref{deltaalpha}, \ref{deltaW}) there is no reason, why such terms should be absent.

We will discuss 
possible corrections to the energies and
to the observable (\ref{EnergiesOdom}, \ref{aFormula}) up to order $\alpha \delta_c^2$ in a systematic way.
We address the issue of the seemingly relevant corrections $\sim \alpha \delta_c$ and $\alpha \delta_W^2$.

Note that there are further dimensionless scales. In particular, if one would consider axial motion with the axial frequency $\nu_z$, there would be a fourth small dimensionless parameter 
\be\label{deltaz}
\delta_z \equiv\frac{h \nu_{z}}{m c^2}\sim 10^{-13}\sim \epsilon^{ 6}.
\ee
This parameter is shortly mentioned in the discussion.
There are, of course, further higher orders in fine structure constant $\alpha$,  which are not shown in figure~\ref{fig:Moti} since these corrections do not introduce new effective operators.

Note further that,
in principle, this discussion of orders of magnitude also applies to the anomalous magnetic moment of the muon,
but there are two crucial differences: First, the experimental observable is different from~(\ref{aFormula}).
Second the muon is about two hundred times heavier than the electron, which further suppresses $\delta_c$ by the same factor.
Thus, most of the possible corrections discussed for the electron are irrelevant for the muon.

\section{Corrections from effective Lagrangian approach}
\label{sec_Quant}

We address the problem of corrections to the anomalous magnetic moment of the electron with an effective Lagrangian approach. 
This approach allows to obtain the energy spectrum as a result of the solutions of the effective equations of motion.
Alternatively the energies can be read off from the spectral structure of the exact electron propagator in the presence of a finite magnetic field, using Schwinger`s proper-time representation~\cite{Ayala:2021lor}. We chose the former approach,
because it allows to include non-QED effects straight forwardly.

\subsection{Hamiltonian from effective Lagrangian}
\label{subsec_Ham}

In an effective field theory approach the anomalous magnetic moment of the electron
is given in terms of the effective Lagrangian~\cite{Itzykson:1980rh}
\be\label{Leff}
{\mathcal{L}}_{eff}=\bar \psi
\left[\gamma^\mu (i\hbar \partial_\mu-e A_\mu )-m c^2+a
\frac{e \hbar}{4m} \sigma^{\mu \nu} F_{\mu \nu}+ \dots\right]\psi.
\ee
The anomalous contribution to the $g$-factor  is encoded in the parameter $(a)$. In a top-down approach the corresponding last
term in (\ref{Leff})
 arises from the sunset diagram with a single external photon line (see figure \ref{fig:Sunset} with $y=1$). 
In a bottom-up approach, one can also 
take $(a)$ as a phenomenological parameter of the effective Lagrangian (\ref{Leff}) and determine its value from an experiment. For this purpose, one uses (\ref{Leff}) to derive
the corresponding 
quantum mechanical Dirac equation 
\be\label{DiracEff}
\left[\gamma^\mu (i\hbar \partial_\mu-e A_\mu) -m c^2+a
\frac{e \hbar}{4m} \sigma^{\mu \nu} F_{\mu \nu}+\dots \right]\psi=0.
\ee
For a constant external magnetic field, and $a=0$ the energy levels and eigenfunctions of the system with a single electron are known exactly~\cite{Thaller:1992ji,Suzuki:2005wra},
with the eigenenergies given in (\ref{QMexact}).
With $a\neq 0$, the anomalous magnetic moment is described by an additional term, 
\be\label{Delta1H}
\Delta_1 H= a \gamma^0 \sigma^{\mu \nu} F_{\mu \nu},
\ee
adding to
the Dirac Hamiltonian. 
Since these
are also eigenfunctions of the operator $\Delta_1 H$ the corresponding energy levels (\ref{QMexact})
only get modified by an additive factor
\be\label{EnExact}
E_n^\pm= m c^2\sqrt{1+ \frac{h \nu_c}{m c^2}(1+2n\pm 1)}\pm \frac{a}{2} h \nu_c.
\ee
Since our aim is to study corrections to~(\ref{EnergiesOdom})
we expand the energies (\ref{EnExact}) up to powers of $\delta_c^3$
\be\label{Endeltac3}
E_n^\pm=mc^2+\frac{1}{2}
\left( 2 n+ 1\pm 1 \right)h \nu_c
\pm \frac{a}{2}h \nu_c
-\frac{1}{8}
\frac{h^2 \nu_c^2}{m c^2} \left( 2 n+ 1\pm 1 \right)^2
+\frac{1}{16}
\frac{h^3 \nu_c^3}{m^2 c^4}
\left( 2 n+ 1\pm 1 \right)^3.
\ee
One notes that this contains the constant mass term, 
the energy formula (\ref{EnergiesOdom}),
and the additional last term $\sim \delta_c^3\sim \epsilon^{12}$,
which is the type of correction we are interested in.
Schematically one can write
\bea\label{ENSchema}
\frac{E_n^\pm}{m c^2 \delta_c}&=&{\mathcal{O}}
\left( \frac{1}{\delta_c} + 1 + \alpha + \delta_c + \delta_c^2+ \dots \right)\\ \nonumber
\eea
The contributions $\sim \delta_c$
lead to relative changes in the energy differences of $\sim 10^{-9}$
and are thus an essential part of the current measurements which have a precision of about $10^{-13}$.
The corrections $\sim \delta_c^2$
lead to relative changes in the energy differences of $\sim 10^{-18}$, they will only be accessible to future experiments.
One realizes that the solution (\ref{EnExact}) does not give rise
to $\alpha \delta_c$ or $\alpha \delta_W^2$ terms. For these terms to appear one needs to consider additional operators in the effective action~(\ref{Leff}).

In the following subsections, we will explore this possibility.

\subsection{Between theory and experiment}
\label{subsec_difference}

There is a conceptual gap between the theoretical definition
of the anomalous magnetic moment and the observable which is used to measure this quantity.
From theory side, one defines $a$ in terms of the photon-electron vertex function $\Gamma_{e \bar e \rho}(p,-p,q)$,  evaluated at zero momentum transfer~\cite{Czarnecki:1995wq,Heinemeyer:2004yq}.
From the experimental side, $a$ is defined through equation~(\ref{aFormula}) in terms energy eigenvalues of an electron in the background of a finite magnetic field.

These are different things, which only agree in certain limits and approximations.
One way to discuss this issue is in terms of the effective Lagrangian (\ref{Leff}) and the resulting field equations, as sketched in the previous subsection.
The photon-electron vertex function $\Gamma_{e \bar e \rho}(p,-p,q)$ evaluated at zero momentum gives rise to
the effective Hamiltonian contribution (\ref{Delta1H}). This contribution is thus an unambiguous SM prediction,
which is experimentally tested by measuring transitions between energy eigenvalues of (\ref{DiracEff}) and plugging them into relation  (\ref{aFormula}).
This test is correct as long as additional contributions to the effective Lagrangian, symbolized by the dots in relations (\ref{Leff}, \ref{DiracEff}) are negligible
or at least give negligible contributions to the energy eigenvalues in  (\ref{aFormula}).

Higher order corrections to the effective field equations (\ref{DiracEff}) which are not included in the standard analysis 
because they do not arise from the photon-electron vertex function $\Gamma_{e \bar e \rho}(p,-p,q)$
are a consequence of three different aspects:
\begin{itemize}
\item The experiments are performed at finite magnetic field, wich means that the assumption of zero momentum transfer
is only approximately true. Since such corrections are typically quadratic in the momentum transfer, one can expect
that they lead to $\sim \delta_c^2$ corrections in (\ref{ENSchema}).
\item When allowing contributions to second or higher order in the finite external magnetic field, (\ref{DiracEff}) will contain
additional terms which have a field and Lorentz structure, which is not captured by (\ref{Delta1H}).
Such corrections can already appear at order $\sim\alpha \delta_c$ in (\ref{ENSchema}).
\item Additional, non QED, fields can also lead to operators with a different structure than (\ref{Delta1H}) and which are thus
not part of the standard analysis in the literature. If the additional field is a weak gauge boson, one expects $\sim\alpha \delta_W^2$
corrections to (\ref{ENSchema}). If the additional field is some hypothetical beyond the Standard Model field, one expects
a correction suppressed by the coupling and by the mass of this field.
\end{itemize}
In the following discussion we will estimate these effects for the Standard Model of particle physics,
but some of the findings can be useful for studies that go beyond this model.

\subsection{Further gauge invariant bilinear corrections}
\label{subsec_further2}

\begin{figure}[ht!]
\begin{center}
\includegraphics[width=0.7\columnwidth]{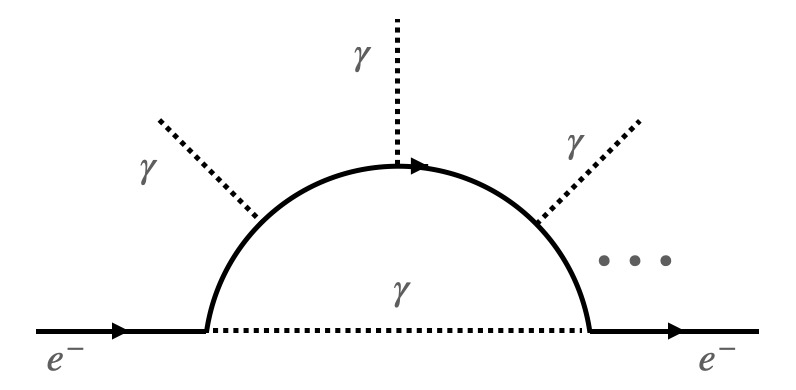} 
\end{center}
\caption{Feynman diagram for
coupling to $y=3$ external photons.}
\label{fig:Sunset}
\end{figure}

In a top-down approach, further corrections to the effective action~(\ref{Leff}) arise from irreducible Feynman diagrams with an arbitrary number of external legs.
We are interested in the cases with two external fermion lines and $y$-external photon lines, as shown in figure~\ref{fig:Sunset}.
For $y=1$, the only contribution is the one given in
 (\ref{Leff}).
Loop corrections induce numerous terms with $y>1$.
In an effective field theory approach
we study a class of possible 
terms in the extended Lagrangian  that meet the following criteria:
\begin{itemize}
    \item they are bi-linear in the Dirac field $\bar \psi {\mathcal{O}} \psi$
    \item they form a Lorentz scalar
    \item the field content of the operator ${\mathcal{O}}={\mathcal{O}}(A_\mu, F^{\mu \nu})$ 
     only contains  gauge-invariant contributions of the external electromagnetic field which are either the field strength tensor $F^{\mu \nu}$, or the covariant derivative operator $D_\mu$, thus 
    ${\mathcal{O}}={\mathcal{O}}(D_\mu, F^{\mu \nu})$.
    \item the field content of ${\mathcal{O}}$ is only up to second power in the field and has at least one contribution from $F^{\mu \nu}$
\end{itemize}
Since the operator ${\mathcal{O}}$ forms a $4\times 4$ matrix in spinor-space,
it is useful to use 
$\{1,\gamma^\mu, \sigma^{\mu \nu}, \gamma^5 \gamma^\mu , \gamma^5\}$ as
complete, five-dimensional basis for the classification of these operators. While the first three basis elements are naturally induced in QED, the last two elements require virtual non-QED contributions such as weak interactions or BSM physics.

\subsubsection{$\sim 1$ contributions}

The term one can add to
the effective Lagrangian (\ref{Leff}), which is proportional to the identity matrix, is
\bea\label{PC1}
\Delta_{1,FF}{\mathcal{L}}&=&\alpha\frac{ \xi_{1,FF}}{m^3 c^6}\; \bar \psi F_{\mu \nu}F^{\mu \nu} \psi,
\eea
where $\xi_i$ is the effective dimensionless coupling parameter, which is expected to be of order one.
The contributions of these operators
to the energy spectrum can be estimated by using $|\phi_n^\pm\rangle$, the  eigenfunctions 
of (\ref{DiracEff}), which are given in (\ref{Eigenfunctions}).
These eigenfunctions are normalized as $\langle \phi_n^\pm|\phi_n^\pm\rangle=1$.
Since small terms like (\ref{PC1}) only lead to small modifications of these wave-functions one can apply perturbation theory to estimate to which extend these modifications 
shift the energy levels $E_n^\pm$.
These shifts are labeled $\Delta_{\dots }E_n^\pm$.
One finds 
\bea\label{DE0}
\Delta_{1,FF} E_n^\pm&=&\frac{ \xi_{1,FF}\alpha}{m^3 c^6}
\langle \phi_n^\pm| 
\gamma^0 F_{\mu \nu}F^{\mu \nu}|\phi_n^\pm\rangle\\ \nonumber
&=& -\xi_{1,FF}\alpha\frac{h^2 \nu^2 }{2m c^2 \sqrt{1+\frac{h \nu}{m c^2}(2n+1\pm 1)}}
\\ 
&\approx& \xi_{1,FF}\alpha\left(-\frac{h^2 \nu_c^2}{m c^2}+ \frac{h^3 \nu_c^3 (2n+1 \pm 1)}{m^2 c^4}\right).
\eea
In terms of the series in figure \ref{fig:Moti}, these contributions are of the order of 
\be\Delta_{1,FF} E_n^\pm/(m c^2 \delta_c)\sim \xi_{1,FF} {\mathcal{O}}(\alpha \delta_c + \alpha \delta_c^2).
\ee

Note that these corrections are
also dependent on the other quantum number $l$, as long as $l\le n$. The above relation is for $l=0$.

Note further that corrections of this type can also appear due to finite temperature effects,
but these are so small that they do not interfere with our discussion.

\subsubsection{$\sim \gamma^\mu$ contributions}

When including one $\gamma^\mu$, the first option is to consider one $D_\alpha$ and one $F^{\mu \nu}$.
Since 
$D_\alpha \gamma^\alpha F^\mu_\mu=0$,
the only possibility, which is linear in the field strength is
\bea\label{PC2}
\Delta_{\gamma,DF}{\mathcal{L}}&=&\frac{ \xi_{\gamma,DF}}{m^2 c^4}\; \bar \psi D_\alpha \gamma^\beta F^\alpha_\beta \psi\neq 0.
\eea
However, when integrating the corresponding contribution to the energy,
one finds that it vanishes
\bea
\Delta_{\gamma,DF} E_n^\pm&=&\frac{ \xi_{1,DF}}{m^2 c^4}
\langle \phi_n^\pm| 
\gamma^0 D_\alpha \gamma^\beta F^\alpha_\beta |\phi_n^\pm\rangle\\ \nonumber
&=& 0.
\eea
Second-order in $F^{\mu \nu}$, 
there are two non-vanishing contributions
\bea\label{PC3}
\Delta_{\gamma,DFF1}{\mathcal{L}}&=&\alpha\frac{ \xi_{\gamma,DFF1}}{m^4 c^8}\; \bar \psi D_\alpha \gamma^\alpha F_{\mu \nu}F^{\mu \nu} \psi
\eea
with
\bea\label{DE1}
\Delta_{\gamma,DFF1} E^\pm_{n}&=&\xi_{\gamma,DFF1}\alpha\frac{h^2 \nu^2 }{m c^2\sqrt{1+h \nu (1+2n\pm 1)}}\\
&\approx &
\xi_{\gamma,DFF1}\alpha \left(\frac{2 h^2 \nu^2}{m c^2}
-\frac{h^3 \nu^3 (1+2n\pm 1)}{m^2 c^4}\right)
\eea
and
\bea\label{PC4}
\Delta_{\gamma,DFF2}{\mathcal{L}}&=&\alpha \frac{ \xi_{\gamma,DFF2}}{m^4 c^8}\; \bar \psi D_\alpha \gamma^\beta F_{\beta \nu}F^{\alpha \nu} \psi
\eea
with
\bea\label{DE2}
\Delta_{\gamma,DFF2}E^\pm_{n}&=&-\Delta_{\gamma,DFF2}\alpha\frac{h^3 \nu^3 (1+2n\pm 1)}{m^2 c^4\sqrt{1+h \nu (1+2n\pm 1)}}\\ 
&\approx &-\xi_{\gamma,DFF2}\alpha\frac{h^3 \nu^3 (1+2n\pm 1)}{m^2 c^4}.
\eea
In terms of the series in figure \ref{fig:Moti}, these contributions are of the order of 
\bea
\Delta_{\gamma,DFF1} E_n^\pm/(m c^2 \delta_c)
&\sim& \xi_{\gamma,DFF1} {\mathcal{O}}(\alpha \delta_c+\alpha \delta_c^2)\\
\Delta_{\gamma,DFF2} E_n^\pm/(m c^2 \delta_c)
&\sim& \xi_{\gamma,DFF2} {\mathcal{O}}(\alpha \delta_c^2).
\eea

\subsubsection{$\sim \sigma^{\mu \nu}$ contributions}

The contribution, which is  first-order in $~F^{\mu \nu}$, is already considered by the coupling in (\ref{DiracEff}).
Second-order contributions such as
$\sim F^{\alpha}_\beta \sigma^{\beta \mu} F_{\mu \alpha}$ are identically zero.
A possible third-order contribution is
\bea\label{PC5}
\Delta_{\sigma,FFF1}{\mathcal{L}}&=&\alpha\frac{ \xi_{\sigma,FFF1}}{m^5 c^{10}}\; \bar \psi \sigma^{\mu \nu} F_{\mu \nu} F_{\alpha \beta }F^{ \alpha \beta} \psi
\eea
with 
\bea\label{DE3}
\Delta_{\sigma,FFF1} E^\pm_n&=&\pm \xi_{\sigma,FFF1}\alpha \frac{4 h^3 \nu^3 }{m^2 c^4}
\eea
Similarly, there is
\bea\label{PC6}
\Delta_{\sigma,FFF2}{\mathcal{L}}&=&\alpha\frac{ \xi_{\sigma,FFF2}}{m^5 c^{10}}\; \bar \psi \sigma^{\mu \nu} F_{\mu \alpha} F_{\nu \beta }F^{ \alpha \beta} \psi
\eea
with 
\bea\label{DE4}
\Delta_{\sigma,FFF2} E^\pm_n&=&\pm \xi_{\sigma,FFF2} \alpha\frac{2 h^3 \nu^3 }{m^2 c^4}.
\eea
Thus, one realizes that in the spirit of 
the expansion in figure \ref{fig:Moti},
all these contributions are of order
\be
\Delta_{\sigma \dots} E_n^\pm/(m c^2 \delta_c)
\sim \xi_{\sigma \dots} {\mathcal{O}}(\alpha \delta_c^2).
\ee
\subsection{Parity violating corrections}
\label{subsec_further3}

Once one leaves the realm of QED, there is also the possibility of forming Lorentz invariants with $\gamma^5, \gamma^\mu \gamma^5$ and through contractions with $\epsilon^{\mu \nu \alpha \beta}$

\subsubsection{$\sim \gamma^5$ contributions}

One might consider mass contributions which have the form 
\bea\label{PV1}
\Delta_{\gamma^5}{\mathcal{L}}&=&m c^2 \xi_{\gamma^5}\alpha \delta_W^2
 \bar \psi \gamma^5 \psi.
\eea
They do not give corrections to the energy 
\bea\label{DE5}
\Delta_{\gamma^5}E_n^\pm&=&0.
\eea

\subsubsection{$\sim \gamma^5 \gamma^\mu$ contributions}

Parity violating contributions to the 
minimal coupling can, for example, be introduced 
by virtual weak interactions 
\bea\label{PV2}
\Delta_{\gamma^5\gamma^\mu}{\mathcal{L}}&=& \xi_{\gamma^5\gamma^\mu}\alpha \delta_W^2
 \bar \psi \gamma^5\gamma^\mu D_\mu \psi.
\eea
Two examples are calculated in the 
appendix~\ref{subsec:PV1}.
Even though these terms come with the expected prefactor,
they do not lead to corrections to the energy levels
\bea\label{DE7}
\Delta_{\gamma^5\gamma^\mu}E_n^\pm&=&0.
\eea
Similarly, there is the possibility of
\bea\label{PV3}
\Delta_{\gamma^5\gamma^\mu FF}{\mathcal{L}}&=&\alpha\frac{ \xi_{\gamma^5\gamma^\mu FF  \delta_W^2}}{m^3 c^6}
 \bar \psi \gamma^5\gamma^\alpha D_\alpha F^{\mu \nu}F_{\mu \nu}\psi,
\eea
also leading to vanishing energy corrections
\bea\label{DE8}
\Delta_{\gamma^5\gamma^\mu FF}E_n^\pm&=&
0.
\eea
Notably, all parity-violating contributions to the energy vanish at the considered order.

\subsubsection{Considering the electric field}

If one considers an electric field,
e.g. the field confining the electrons in $z$-direction, the solutions of equation (\ref{DiracEff}) will contain additional quantum numbers, like $n_z$ referring to this degree of freedom.
\be
\psi = \psi^\pm_{n,n_z, l, l_z} \quad.
\ee
The strength of the electric field will depend on $z$, and it will be related to the dimensionless parameter (\ref{deltaz}) and the energy contribution
of the electric oscillations will be
\be
E_{n,n_z}^\pm-E_{n,n_z=0}^\pm= n_z m c^2 \delta_z+ {\mathcal{O}}(\delta_z^2).
\ee
In order to avoid such complications, 
let us consider a constant electric field 
$\vec E= \hat z (m c^2)^2 \delta_z$.
Thus, as long as we consider only eigenstates with the same quantum number $n_z$ (e.g. $n_z=0$), we can reuse the eigenstates of the previous sections
\be\label{nz0}
\psi^\pm_{n, l}=\psi^\pm_{n,n_z=0, l, l_z=0}.
\ee
The discussion of the preceding sections remains valid in a perturbative sense, even though the electromagnetic stress tensor now reads
\be
F_{\mu \nu}=(m c^2)^2\left(
\begin{array}{cccc}
     0&0&0& \delta_z \\
     0&0&\delta_c &0\\
     0&-\delta_c&0&0\\
     -\delta_z&0&0&0
\end{array}
\right).
\ee
The existence of the electric field $\sim (m c^2)^2 \delta_z$ does not alter the differences between energy levels of the state (\ref{nz0}). This invisibility becomes spoiled 
due to parity-violating contributions, analogous to the ones discussed above. One obtains, for example, an effective coupling
\be\label{PV7}
\Delta_{\sigma \tilde F} {\mathcal{L}}= \alpha \frac{\xi_{\sigma \tilde F} \delta_W^2}{m c^2} \sigma^{\mu \nu} \tilde F_{\mu \nu},
\ee
where the dual electromagentic tensor is given by $ \tilde F_{\mu \nu}= \frac{1}{2} \epsilon_{\mu \nu \alpha \beta} F^{\alpha \beta}$.
This coupling contributes to the energy levels with
\be
\Delta_{\sigma \tilde F} E_n^\pm=\pm m c^2 \alpha \delta_W^2\xi_{\sigma \tilde F}\delta_z.
\ee

\subsection{Combined corrections}
\label{subsec_combinedCorr}

As a result of the previous two subsections, we have obtained eight corrections (\ref{DE0}, \ref{DE1}, \ref{DE2}, \ref{DE3}, \ref{DE4}, \ref{DE5}, \ref{DE7}, \ref{DE8})
of the standard
energy formula (\ref{EnergiesOdom}).
Up to order $\delta_c^3$ the expansion of the standard
energy formula reads  (\ref{Endeltac3}).
Within the same expansion, the combined correction arising from
(\ref{DE0}, \ref{DE1}, \ref{DE2}, \ref{DE3}, \ref{DE4}, \ref{DE5}, \ref{DE7}, \ref{DE8}) 
reads
\bea\label{DeltaECombined}
\frac{\Delta E_n^\pm}{m c^2 \delta_c}&=&  \alpha \delta_c^1 (\xi_{\gamma^\mu FF1}-\xi_{1,FF})\pm
\alpha \frac{\delta_W^2 \delta_z}{ \delta_c}\xi_{\sigma \tilde F}
\\ \nonumber&&
+\alpha \delta_c^2
(
\xi_{Id FF}(2 n \pm 1 +1)
-2n (\xi_{\gamma^\mu FF1}+\xi_{\gamma^\mu FF2})-
\xi_{\gamma^\mu FF1}-\xi_{\gamma^\mu FF2}\\ \nonumber 
&& \quad \quad
\mp (\xi_{\gamma^\mu FF1}+\xi_{\gamma^\mu FF1}-2(\xi_{\sigma FFF1}+\xi_{\sigma FFF2}))
)
+ \dots
\eea
In equation (\ref{xiPred}) a theoretical prediction for the proportionality factor of the leading coefficient of (\ref{DeltaECombined}) is given $(\xi_{\gamma^\mu FF1}-\xi_{1,FF})\approx 0.02$.

\subsection{Anomalous magnetic moment}
\label{subsec_gm2}

Experimentally, the value of $a_d$ is 
obtained from measuring transition energies and applying relation (\ref{aFormula}). 
This 
formula is valid for energies expanded up to $\delta_c^2$.
The value inferred from the direct Penning trap experiment by using this relation is~\cite{Hanneke:2008tm}
\be\label{ad}
a_d=0.001\, 159\, 652\, 180\, 73\,(28).
\ee
If one wants to include the relativistic $\delta_c^3$ energy corrections (\ref{Endeltac3}) one has to adopt the formula (\ref{aFormula}) to 
\be\label{aFormuladeltac3}
a_d=\frac{E_0^+-E_1^-}{E_1^+-E_0+ \frac{3}{2}m c^2 \delta_c-\frac{7}{2}m c^2 \delta_c^2}.
\ee
If one includes now, on top of the relativistic improvement (\ref{Endeltac3}), the combined corrections (\ref{DeltaECombined}), the actual energies read
\be
\tilde E_n^\pm=E_n^\pm+\Delta E_n^\pm.
\ee
Using these energies in the relativistically improved formula (\ref{aFormuladeltac3}) one finds 
the inferred direct anomalous moment $a_d$ (the l.h.s. of (\ref{aFormuladeltac3})) differs from the actual magnetic moment $a_a$ (the one appearing in (\ref{Leff})). 
One finds
\bea\label{adactual}
a_d&=& a_a+ \frac{2 \alpha \delta_W^2 \delta_z}{\delta_c}\xi_{\sigma \tilde F}
-2 \alpha \delta_c^2 \left(a_a(\xi_{Id FF}-\xi_{\gamma^\mu FF1}-\xi_{\gamma^\mu FF2})-2(2 \xi_{\sigma FFF1}+ \xi_{\sigma FFF2})\right)+ {\mathcal{O}}(\epsilon^{11}).
\eea
The above relation used a combined expansion in the small parameter $\epsilon$
(\ref{deltac}, \ref{deltaalpha}, \ref{deltaW}).
This relation can be 
inverted to give
the actual value of the anomalous magnetic moment $a_a$ as a function of $a_a=a_a(a_d,\xi_i)$
\bea\label{aaactual}
a_a&=& a_d-\frac{2 \alpha \delta_W^2 \delta_z}{\delta_c}\xi_{\sigma \tilde F}
+2 \alpha \delta_c^2 \left(a_d(\xi_{Id FF}-\xi_{\gamma^\mu FF1}-\xi_{\gamma^\mu FF2})-2(2 \xi_{\sigma FFF1}+ \xi_{\sigma FFF2})\right).
\eea

It is now the task of theory to predict the precise values of $\xi_i$ and the task of experiment to determine and/or constrain these values. One way of doing this is via a comparison with indirect $g-2$ measurements.

\section{Discussion}
\label{sec_Discussion}

\subsection{Indirect $g-2$ measurements}

Theoretical calculations determine a relation between the value of $(g-2)$ and the fine structure constant up to fifth-order in $\alpha$
\bea\label{aalpha}
a(\alpha)&=& C_2 \left( \frac{\alpha}{\pi}\right)+
C_4 \left( \frac{\alpha}{\pi}\right)^2+C_6 \left( \frac{\alpha}{\pi}\right)^3+C_8 \left( \frac{\alpha}{\pi}\right)^4
\\ \nonumber &&+C_{10} \left( \frac{\alpha}{\pi}\right)^5+a_{\mu \tau} + a_{had}+ a_{weak},
\eea
where the constants $C_i$ are given in~\cite{Gabrielse:2006gg} (consider the erratum).
Thus, an independent measurement of $\alpha$ (and the lepton masses), also allows to infer a value for $g-2$, and vice versa.

\begin{figure}[ht!]
\begin{center}
\includegraphics[width=0.6\columnwidth]{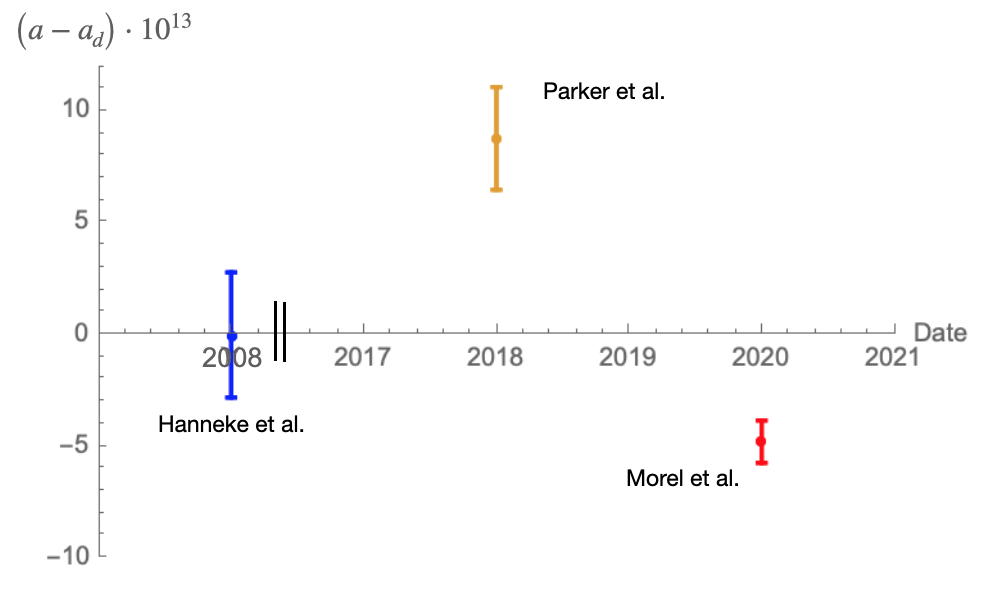}
\end{center}
\caption{
Results for anomalous magnetic moment $(a_{P/M}(\alpha)-a_d)\cdot 10^{13}$ plotted as a function of time. 
The blue point is the direct measurement~\cite{Hanneke:2008tm}, the red and orange points are the indirect measurements~\cite{Parker:2018vye,Morel:2020dww}.
}
\label{fig:aTime}
\end{figure}

In recent years this has been done by Parker et al. \cite{Parker:2018vye} and Morel et al. \cite{Morel:2020dww}.
They obtained $\alpha$-induced values
\bea\label{aalphaNP}
a_P(\alpha)&=&0.001\, 159\, 652\, 181\, 610 \,(230)\\ \label{aalphaNM}
a_M(\alpha)&=&0.001\, 159\, 652\, 180\, 252 \,(95).
\eea
In contrast to this, the ``direct'' measurement gave~(\ref{ad}).
 This situation is shown in figure \ref{fig:aTime}, where we plot the value of $a(\alpha)$ relative to the value $a_d$, which was inferred from the direct measurement. 
Since the latter measurement of $\alpha$ has much smaller error bars, there seems to be no significant tension between $a_d$
and $a_M(\alpha)$. 
The (non)-tension between 
the direct measurement and the indirect measurements can be quantified by
\be
 \frac{a_{P/M}(\alpha)-a_d}{\sqrt{\delta a_d^2+\delta a^2_{P/M}(\alpha)} }= 
\left\{
\begin{array}{cc}
    t_{i,P} = +2.4\\
    t_{i,M} = - 1.6
\end{array}
\right. .
\ee
The values of $t_{i,P/M}$ are measured in standard deviations.
 Here the sign indicates whether $a(\alpha)$ is smaller or bigger than $a_d$.

In the following subsections, we will explore a comparison between $a_d$ and $a(\alpha)$ in the light of the difference between $a_d$ and $a_a$ (\ref{aaactual}).
In this comparison we will identify $a(\alpha)$ with $a_a$. This identification assumes implicitly 
that there are no corrections to the 
 measurements of the fine-structure constant $\alpha$.

\subsection{Benchmark without parity violation}

Let's first consider a benchmark without parity violation by setting $\xi_{\sigma \tilde F}=0$.
The result in section \ref{sec_Quant} suggests correcting
the interpretation of the direct 
measurement by using the value 
(\ref{adactual}): $a_d
\rightarrow a_a=a_a(\xi_i)$
and the error induced by this correction.

\begin{figure}[ht!]
\begin{center}
\includegraphics[width=0.6\columnwidth]{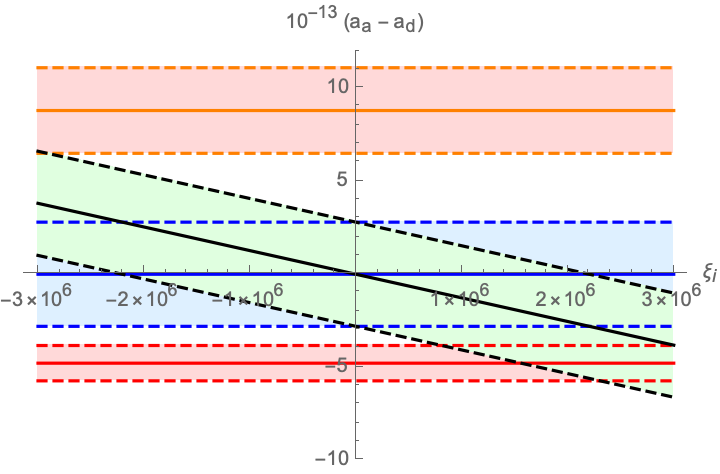}
\end{center}
\caption{
The red line shows
$(a_M-a_d)$,
with the error bars of $a_M$~\cite{Morel:2020dww}.
The orange line shows
$(a_P-a_d)$,
with the error bars of $a_P$~\cite{Parker:2018vye}. The blue region are the error bars of $a_d$~\cite{Hanneke:2008tm}. The black line shows
$(a_a(\xi_i)-a_d)$,
with the error bars of $a_d$.
}
\label{fig:axi}
\end{figure}

As the first benchmark, let's treat
all parity conserving contributions with the same numerical factor $\xi_i$.
For this benchmark, one can study how
the comparison between direct measurement and
indirect measurement depends on the parameter $\xi_i$. 
If one assumes these corrections come without an increased error bar, one can plot $a_a(\xi_i)$ relative to $a_d$ and in comparison with the indirect results $a_{P/M}-a_d$.
This is shown in figure~\ref{fig:axi}.

From figure \ref{fig:axi} one sees that the parity conserving corrections
only have a relevant effect on the comparison 
between direct and indirect observation, if the dimensionless parameters $\xi_i$ are of the order of $10^6$.
However, the prefactors of the effective operators are chosen such that within the SM the parameters $\xi_{i,\dots}$ are expected to be of order one.
Thus, figure \ref{fig:axi} means that there are six orders of magnitude missing in the sensibility before one can appreciate the effect of these SM contributions.

This finding can be compared with the QED corrections for finite momentum transfer.
The one-loop QED form factor for finite
photon momentum squared $q^2$ is given by
\bea\label{QFT1}
F_2(q^2)&=& \frac{\alpha}{2 \pi}\int_0^1 dx dy dz \;\delta(x+y+z-1)\frac{2 m^2z (1-z)}{m^2(1-z)^2-q^2 x y}\\ \nonumber
&=&\frac{\alpha}{2 \pi}\left(1- \frac{q^2}{6 m^2}\right).
\eea
If one evaluates the correction due to the second term with the momentum scale given by the magnetic background field $q^2\approx (h \nu_c)^2$, one obtains
\be
F_2\approx \frac{\alpha}{2 \pi}\left(1-\frac{\delta_c^2}{6}\right).
\ee
Thus, one also finds that finite momentum effects are suppressed by $\alpha \delta_c^2\approx 10\cdot 10^{-20}$, which is also
six orders of magnitude below the current experimental precision.

From a phenomenological perspective (e.g. for BSM models) one has to demand that $|a_a-a_d|\le 10\cdot 10^{-13}$ which constrains the $\xi_i$ parameter of this benchmark to
\be
\xi_i\le 1\cdot 10^{6}.
\ee
The parameters $\xi_i$ are thus useful mediators between BSM models and the precision measurement of $a_d$.

\subsection{Benchmark with parity violation}

\begin{figure}[ht!]
\begin{center}
\includegraphics[width=0.6\columnwidth]{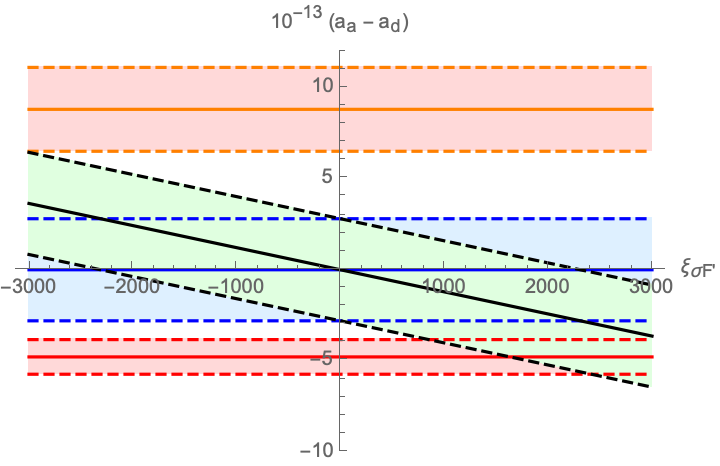}
\end{center}
\caption{
The red line shows
$(a_M-a_d)$,
with the error bars of $a_M$~\cite{Morel:2020dww}.
The orange line shows
$(a_P-a_d)$,
with the error bars of $a_P$~\cite{Parker:2018vye}. The blue region are the error bars of $a_d$~\cite{Hanneke:2008tm}. The black line shows
$(a_a(\xi_i)-a_d)$,
with the error bars of $a_d$.
}
\label{fig:axiPV}
\end{figure}

In a second benchmark, we set all parity conserving corrections to zero and only retain $\xi_{\sigma \tilde F}$.
Figure \ref{fig:axiPV} shows how the comparison between direct and indirect measurements is affected by the remaining parameter.
One notes that a natural $\xi_{\sigma \tilde F}={\mathcal{O}}(1)$ is three
orders of magnitude too small for observation under the current simplifying assumptions.

\subsection{What happened?}

In subsection \ref{subsec_expect} it has been shown that background field effects $(\sim \alpha \delta_c)$
and effects of the weak interaction $(\sim \alpha \delta_W^2)$ can be expected to give important contributions to the energy levels $E_n^\pm/(m c^2 \delta_c)$, as  depicted in figure \ref{fig:Moti}. These energy levels are used to determine  $a_d$.
However, when studying the actual comparison between direct and indirect measurements, the effect of these corrections was rather negligible, as shown in figure~\ref{fig:axi}.
The disappearance of
these seemingly substantial contributions happens for different reasons:
\begin{itemize}
    \item The parity-violating contributions, arising from virtual weak interactions, introduce new types of effective couplings (\ref{PV1}, \ref{PV2}, \ref{PV3}). These couplings do, however, not contribute to the leading-order corrections of the energy levels $E_n^\pm/(m c^2 \delta_c)$. This is due to the particular off-diagonal nature of the $\gamma^0\cdot \gamma^5$ matrix.
\item 
    The parity conserving corrections (\ref{PC1}, \ref{PC2},\ref{PC3}, \ref{PC4}, \ref{PC5}, \ref{PC6}) do all contribute to $E_n^\pm/(m c^2 \delta_c)$ and they do so with corrections $\sim \alpha \delta_c^1$ and  $\sim \alpha \delta_c^2$. The problem is that all the the $\sim \alpha \delta_c^1$ corrections are independent of the quantum numbers $(\pm, \; n)$. Thus, when evaluating energy differences in (\ref{aFormuladeltac3}), only the much smaller $\sim \alpha\delta_c^2$ corrections survive. 
    It remains to be seen whether larger corrections $\sim \alpha \delta_c^1$ can be noted by a different type of experimental setup.
\item 
    The largest corrections arise from the parity-violating contributions (\ref{PV7}). Under the simplifying  assumptions used above, they are still three orders of magnitude below the sensibility threshold. Concering this result, one has to consider two additional aspects. On the on hand, one could imagine an experimental setup with an increased electric field, which would then reach the observability limit. On the other hand, one must consider that the assumption of a homogenous and constant electric field is not realistic. In a more realistic scenario, it is likely that contributions from positive and negative electric fields cancel each other. 
\end{itemize}

\section{Conclusion}
\label{sec_Concl}

Penning trap experiments allow
for a  precise direct measurement of the electron's anomalous magnetic moment $a=(g-2)/2$. This is done by
relating observed transition energies between
Landau levels, as shown in equation~(\ref{aFormula}). 

In this paper, we studied the impact of parametrized corrections to the effective Lagrangian (\ref{Leff}) on 
the electron's anomalous g factor,
deduced from the observable (\ref{aFormuladeltac3}).
It is found that the presence of these parameters leads to a difference between the deduced anomalous magnetic moment $a_d$ and the actual magnetic moment $a_a$ given by (\ref{adactual}).
This generic result is then studied
for two benchmarks.

 It is shown that there are QED-loop-induced parity conserving corrections to the energies $E_n^\pm$ which are of the order of $\alpha \delta_c^2$. Such corrections are, in principle, numerically very relevant since
    $\Delta a\sim \Delta E/\delta_c\sim \alpha \delta_c \sim 10^{-11}$.
    However,
    these leading corrections
    do not depend on the quantum numbers $(n,\,\pm)$ and thus they get canceled in the actual observable~(\ref{aFormuladeltac3}), where only energy differences are considered. The remaining sub-leading effects $\Delta a_d\sim \delta_c^2$ are,
    at the current stage, not observable.
They are about six orders of magnitude smaller than the experimental precision.

 Most parity-violating corrections to the energy vanish unless one considers effects of the electric field. 
An example for such a contribution is given in relation~(\ref{PV7}).
Whether this scenario has the potential of studying parity violation with $g-2$ experiments has to be investigated with a
detailed study considering a spatial variation of the electric field.

While this paper has its focus on SM contributions, the relations (\ref{DeltaECombined})
are perfectely suited 
to test and constrain BSM effects.

\section*{Acknowledgements}

We thank R. Schoeffbeck, J. Suzuki, A. Rebhan, and F.~Hummel for their comments  
and C.~Kat for his efforts to improve our pronunciation of 
``Foldy-Wouthuysen''.

\appendix
\section{Appendix}

\subsection{ Explicit solutions}

The exact solutions of (\ref{DiracEff}) are well known~\cite{Suzuki:2005wra}.
Since we made massive use of the particle eigenstates $\psi_{n=(0,1), \, l=0}^{\pm}$, these functions in symmetric gauge are given explicitly
\bea\label{Eigenfunctions}
\psi_{0,0}^{+}&=&\left(\frac{\rho}{2 \sqrt{2\pi}} \sqrt{1+\frac{m}{\sqrt{m^2+\rho^2}}},0,0,
\frac{i \rho^3 (x+iy)}{4 \sqrt{2\pi} \sqrt{m(\sqrt{m^2+\rho^2}+m)+\rho^2}} \right)e^{-\frac{1}{8} \rho^2(x^2+y^2)-it \sqrt{m^2+\rho^2}}\\ \nonumber
\psi_{0,0}^{-}&=&\left(0,\frac{\rho}{2\sqrt{\pi}},0,0 \right)e^{-\frac{1}{8} \rho^2(x^2+y^2)-itm }
\\ \nonumber
\psi_{1,0}^{+}&=&
\left(\frac{\rho^2(x+iy)}{4 \sqrt{2 \pi}} \sqrt{1+\frac{m}{\sqrt{m^2+\rho^2}}},0,0,
\frac{i \rho^4 (x+iy)^2}{8 \sqrt{2\pi} \sqrt{m(\sqrt{m^2+2\rho^2}+m)+2\rho^2}} \right)e^{-\frac{1}{8} \rho^2(x^2+y^2)-it \sqrt{m^2+ 2\rho^2}}
\\ \nonumber
\psi_{1,0}^{-}&=&
\left(\frac{ \rho^2(\sqrt{m^2 +\rho^2}+m)(x+iy)}{4 \sqrt{2\pi} \sqrt{m(\sqrt{m^2+2\rho^2}+m)+2\rho^2}},0,0\frac{-i \rho^2}{2 \sqrt{2\pi} \sqrt{m(\sqrt{m^2+2\rho^2}+m)+2\rho^2}}\right)
e^{-\frac{1}{8} \rho^2 (x^2+y^2)-it \sqrt{m^2+ \rho^2}},
\eea
where the abbreviation $\rho^2=2 m h \nu$ and $c\equiv1$ was used.

\subsection{QED background corrections from Suszuki approach}

In the effective Lagrangian approach (\ref{Leff}), $\xi_{i}$ are phenomenological parameters that have to be determined by experiment. In a full QFT approach, with non-vanishing background fields, these parameters can be predicted. Below we will sketch how such a calculation relates to the above results.


Complementary to using the effective action (\ref{Leff}), one can also calculate the loop corrections to the different energies $E_n^\pm$ directly in the underlying quantum field theory. 
Usually, this is done to leading order in the magnetic field.
For a comparison with the results obtained in the previous sections,
this has to be done up to second order in $B$.
While the complete calculation of these QED corrections goes beyond the scope of this paper, one can use the known shift 
in the ground state energy to get an estimate
of the uncertainty.
The finite ground state corrections, which are proportional to $B^2$, are given by Suzuki~\cite{Suzuki:2005wra}. 
Strong field effects in a broader sense are
discussed in~\cite{Dunne:2008kc}.

Suzuki uses the exact QM solution for the energy levels of the electron~\cite{Suzuki:2005wra} 
\be\label{QMexact}
E_n^\pm= m c^2\sqrt{1+ \frac{h \nu_c}{m c^2}(1+2n\pm 1)},
\ee
Note that when comparing the above formula with the energies given in \cite{Itzykson:1980rh,Suzuki:2005wra}
one needs to do the identification $n\rightarrow n+1$. The
Energies \eqref{QMexact} are identical to~(\ref{EnExact}) for $a=0$. 

The actual (QFT corrected) energies are then labeled ${\mathcal{E}}_n^\pm$, they can 
be expressed as corrections to (\ref{QMexact}). One finds
\be
\Delta E_n^\pm\equiv
{\mathcal{E}}_n^\pm - E_n^\pm = \int d^3x \int d^3x' \bar \psi_{E,n,\pm}(x) {\mathcal{M}}(E_n^\pm,\vec x, \vec x')\psi_{E,n,\pm}(x'),
\ee
where ${\mathcal{M}}$ is the mass operator and $\psi_{E,n,\pm}$ are the eigenfunctions  for the QM eigenvalues (\ref{QMexact}).
The calculation is tedious, but for a small magnetic field
$h \nu_c\ll m c^2$, the result can be written as an expansion
\be\label{DeltaEn}
\Delta E_n^\pm =
\frac{a}{2}h \nu_c\left( 
1+\frac{h \nu_c}{m c^2}(b_0^\pm + b_1^\pm n )
\right)
+ {\mathcal{O}}\left(\frac{h \nu_c}{m c^2} \right)^3.
\ee
In the recent paper of Kim at al.~\cite{Kim:2021ydg}, this type of expansion is given for the ground state energy and the coefficients are calculated explicitly.
However, the above formula allows also for $n\neq 0$.
These energy contributions do, in principle, depend on the quantum numbers $(n,\pm)$, since the eigenfunctions $\psi_{n,\pm}$ depend on these quantum numbers. This dependence on $n$ does not have to be linear, but since (\ref{aFormula}) only considers $n=(0,1)$ a distinction with different powers of $n$ is not necessary.
Here, $a,b_i^\pm$ are numerical coefficients that can be obtained in a loop expansion
\bea
a&=& \frac{\alpha}{2 \pi} + {\mathcal{O}}(\alpha^2)\\
b_i^\pm&= &b_{i,0}^\pm+\alpha \cdot b_{i,1}^\pm+{\mathcal{O}}(\alpha^2).\label{bpmi}
\eea
The coefficients of $a$ are very well studied to high order, while the coefficients $b_i^\pm$ are hard to find in the literature.
Their order of magnitude can be estimated from 
the $\sim B^2$ correction to the ground state energy given in~\cite{Suzuki:2005wra}.
Using the leading order expansion in $\epsilon$
and minimal subtraction, this ground state correction is
\be\label{DeltaE0}
\Delta E_0^-|_{B^2}=\alpha B^2 \frac{1259 }{16800 }
\frac{e^2  \hbar^2}{m^3 c^2\pi}
+{\mathcal{O}}\left(\epsilon^8\right).
\ee
From this, one can read off the numerical value of the first coefficient of (\ref{bpmi})
\be\label{b0m}
b_{0}^-=\frac{1159}{4200}\approx 0.3+ {\mathcal{O}}(\alpha).
\ee
From equation (\ref{DeltaECombined}) and the preceding discussion, one can expect that the
coefficients which are proportional to $n$, vanish at order $\delta_c$ $b_{b_1}^\pm=0$.
Further, one can extract from a comparison between (\ref{b0m}) and (\ref{DeltaECombined}) a theoretical prediction for a combination of the parameters $\xi_{\gamma^\mu FF1}$ and $\xi_{id FF}$
\be\label{xiPred}
\xi_{\gamma^\mu FF1}-\xi_{1,FF}=\frac{b_0-}{4 \pi}\approx 0.02.
\ee

\subsection{Parity violating SM processes}
\label{subsec:PV1}

There are,  SM processes that can contribute to parity violation in low energy QED processes, see for example
\cite{Dittmaier:1997dx,Jones:2004yd,Zykunov:2005md,Du:2019evk}.

\begin{figure}[ht!]
\begin{center}
\includegraphics[width=0.4\columnwidth]{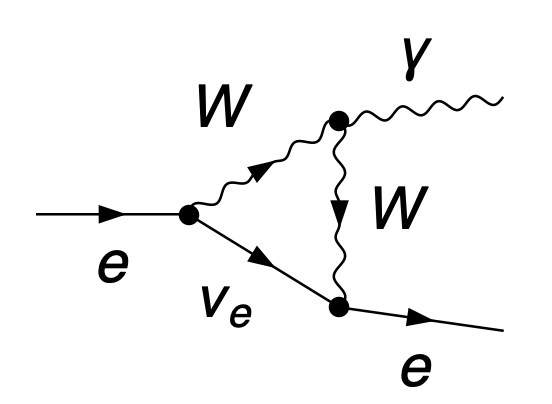} 
\end{center}
\caption{Feynman diagram 
weak sub-process}
\label{fig:Weak1}
\end{figure}

To quantify the magnitude
of similar effects in the context of the 
observable (\ref{aFormuladeltac3})
one needs to consider the weak process shown in figure~\ref{fig:Weak1}.
The prefactor of the Feynman amplitude of this process gives the magnitude 
one should expect for the parameters $\xi_{\gamma^5\dots}$.
As discussed in the previous section,
the contribution, relevant for the $g-2$ observable is $\xi_{\gamma^5 \gamma^\mu}$.
The amplitude shown in figure \ref{fig:Weak1} can be calculated with the software packages based on
\cite{Denner:1991kt,Hahn:2000kx,Hahn:2016ebn}.
It reads
\bea
A&=&\frac{2 i \pi^3 \alpha e}{\sin^2(\theta_W)}
\left(
-2(-2 B_0(m^2,0,m_W^2)+B_0(0,m_W^2,m_W^2)
-2m_W^2 C_0(0,m^2,m^2,m_W^2,m_W^2,0)\right.\\ \nonumber
&&\left.
+2m^2 C_0(0,m^2,m^2,m_W^2,mW^2,0)
-D C_{00}(m^2,0,m^2,0,m_W^2,m_W^2))
+2 C_{00}(m^2,0,m^2,0,m_W^2,m_W^2)\right.\\ \nonumber
&&\left.
+ 3 m^2 C_1(m^2,0,m^2,0,m_W^2,m_W^2)
\right. \cdot \bar \psi_2 \gamma^\mu \gamma^7 \psi_1\\ \nonumber
&&m \bar \psi_2 \gamma^7 \psi_1 \cdot (D-2)p_2^\mu ((D-8)C_1(m^2,0,m^2,0,m_W^2,m_W^2)
2(D-2)C_{12}(m^2,0,m^2,0,m_W^2,m_W^2)))\\ \nonumber
&&m \bar \psi_2 \gamma^6 \psi_1 \cdot
((D-2)p_1^\mu(C_1(m^2,0,m^2,0,m_W^2,m_W^2)
+2 C_{11}(m^2,0,m^2,0,m_W^2,m_W^2))\\ \nonumber
&& p_2^\mu ((D-8)C_1(m^2,0,m^2,0,m_W^2,m_W^2)
+2(D-2) C_{12}(m^2,0,m^20,m_W^2,m_W^2)))\\ \nonumber
&&
+6m^2 C_1(m^2,0,m^2,0,m_W^2,m_W^2)\cdot
\bar \psi_2 \gamma^\mu \gamma^6 \psi_1),
\eea
where $D=4-\tilde \epsilon$ is the spacetime dimension of dimensional regularization, $\gamma^6=(1+\gamma^5)/2$, $\gamma^7=(1-\gamma^5)/2$, and $C_{ij}$ are the Passarino-Veltman loop integral functions which are evaluated with \cite{Denner:2005nn,Patel:2015tea}.
To reduce the resulting expression to the parity-violating
terms, which are relevant for our analysis, we use the following simplifications
\begin{itemize}
    \item retain only terms proportional to 
    $\bar \psi_2 \gamma^\mu \gamma^5 \psi_1$
    \item subtract divergent terms $\sim 1/\tilde \epsilon$
    \item perform an expansion in $\delta_W^2$ in which $A\sim \delta_W^0+ \delta_W^2+ \delta_W^4$. Since the order $\delta_W^0$ is controlled by the dimensional regularization parameters ($\tilde \epsilon$, $\mu$) and since the order  $\delta^4_W$ is negligible, we keep
     the $\sim \delta_W^2$ terms
    \item set $D\rightarrow 4$.
\end{itemize}
The result is then
\be\label{AW}
A|_{WW}=
- i \frac{5 e^3 m^2}{96 \pi^2 m_W^2 \sin^2(\theta_W)} \bar \psi_2 \gamma^\mu \gamma^5 \psi_1.
\ee
The same procedure can be repeated with the
triangle diagram with a single Z-boson in the loop giving
\be\label{AZ}
A|_{Z}=
 i \frac{ e^3 m^2 (6 \log (m_Z^2/m^2)-1)(4 \sin^2(\theta_W)-1)}{192 \pi^2 m_Z^2 \sin^2(\theta_W)\cos^2(\theta_W)} \bar \psi_2 \gamma^\mu \gamma^5 \psi_1.
\ee

Thus one can conclude that parity-violating corrections to the QED couplings are indeed suppressed by 
\be\label{WeakSup}
\sim \alpha \frac{m^2}{m_{W/Z}^2}.
\ee


\end{document}